\documentstyle[11pt,newpasp,twoside,epsf]{article}
\def\edcomment#1{\iffalse\marginpar{\raggedright\sl#1\/}\else\relax\fi}
\reversemarginpar

\def\kms{\ifmmode {\rm km\ s}^{-1} \else km s$^{-1}$\fi}
\def\civ{\ifmmode {\rm C}\,{\sc iv} \else C\,{\sc iv}\fi}
\newcommand \et {{et~al.\ }} 
\newcommand{\half}{\mbox{$\frac{1}{2}$}}
\newcommand{\gsim}{\stackrel{\scriptscriptstyle >}{\scriptstyle {}_\sim}}

\begin{document}
\title{Narrow C\,{\small \bf IV} $\lambda$1549\AA\ Absorption Lines in Moderate-Redshift Quasars}
 \author{M. Vestergaard}
\affil{Department of Astronomy, The Ohio State University, 140 West 18th Avenue, Columbus, OH 43210}

\begin{abstract}
A large, high-quality spectral data base of well-selected, moderate-redshift 
quasars is used to characterize the incidence of narrow associated \civ\ $\lambda$1549 
absorption, and how this may depend on some quasar properties. 
Preliminary results of this study are presented.
\end{abstract}

\section{Introduction}
\vspace{-0.2cm}
Associated narrow absorption lines (NALs) in the spectra of active galaxies 
have widths less than a few hundred km\,s$^{-1}$ and are located within 
5000\,km\,s$^{-1}$ of the emission redshift (Weymann \et 1979).  They are 
likely physically connected to the active galactic nucleus. The working 
hypothesis in this study is that NALs are possibly the low-velocity 
equivalents to the more dramatic broad absorption features (BALs), with line 
widths reaching tens of thousands of km\,s$^{-1}$.  The current, commonly adopted
physical interpretation is that the line widths of both NALs and BALs trace the 
outflow velocity of the absorbing gas and that these outflows are somewhat 
equatorial. The exact solid angle extension of the NAL and BAL matter above 
the disk is unknown.
BALs are predominantly found in the high-luminosity, radio-quiet quasars 
with a frequency of 10--12\%. NALs appear present in 50--70\% of the 
low-luminosity Seyfert galaxies (Hamann 2000), yet the frequency in quasars and how it 
may depend on source radio power and source axis inclination are not 
accurately known.  The aim of this study is to address this issue with 
a large, high-quality, UV spectral data base of $z\approx$\,2 radio-loud
and radio-quiet quasars (hereafter RLQs and RQQs, respectively)
for which the data are uniformly processed and analyzed (Vestergaard 2000; 
Vestergaard \et 2001). The frequency of associated C\,{\sc iv} NALs is 
studied and possible trends with quasar properties, such as luminosity, 
UV spectral slope, radio loudness, and source inclination are tested for.
In particular, if there is a wind evaporating off the accretion disk,
then one might na\"{\i}vely expect this wind to be stronger in brighter, bluer
objects, as the stronger radiation field will blow off more disk matter.
If the associated NALs are somehow related to such a wind one would 
then expect a relationship between the strength of the NALs and the 
continuum characteristics: the continuum luminosity, L$_{\rm cont}$, 
the UV continuum slope, $\alpha_{UV}$, and/or the absolute magnitude, 
M$_{\rm V}$, of the object.
\begin{figure}[t]
\plotfiddle{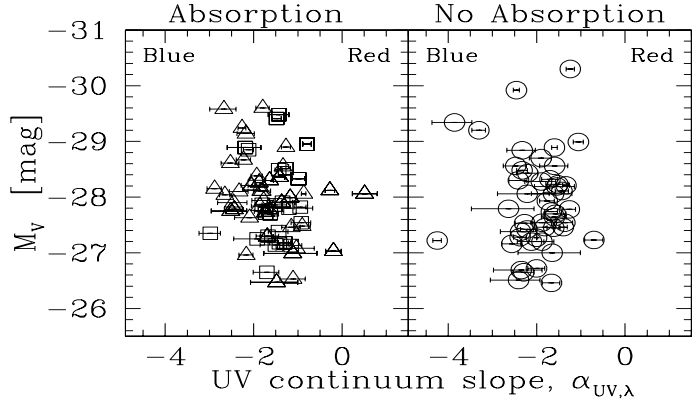}{4.55cm}{-90}{35}{30}{-140}{161}
\caption{The distribution of M$_{\rm V}$ and $\alpha_{\rm UV,\lambda}$ 
for the quasars with (triangles: RLQs; squares: RQQs) and without 
\civ\ NALs (circles).}
\end{figure}

\begin{figure}
\plotfiddle{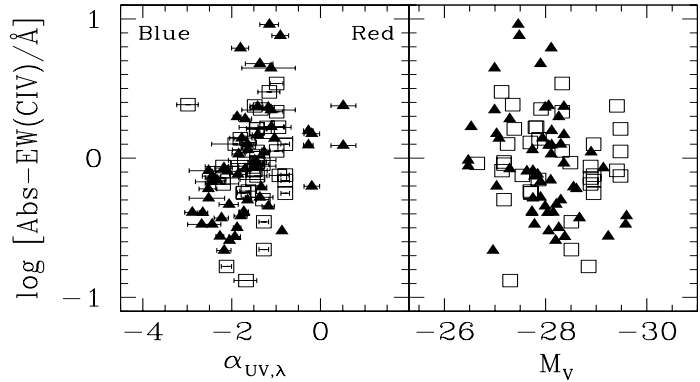}{4.4cm}{-90}{35}{30}{-140}{162}
\caption{The distribution of the strength of the absorber relative to the
UV continuum slope and the quasar luminosity.}
\end{figure}

The first results of this study are presented here. An extended analysis, 
including a detailed comparison of the NAL properties of the RLQs
and the RQQs, will be presented by Vestergaard (2001). 
First, the conclusions are summarized, then the data, measurements, and 
results are presented and briefly discussed.
\section{Summary and Conclusions}
\vspace{-0.15cm}
The main results of this study are as follows:
\begin{itemize}
\vspace{-0.1cm}
\item
The frequency of high-velocity NALs in RQQs and RLQs is similar.
\vspace{-0.2cm}
\item
RLQs show a small excess of associated NALs with respect to RQQs.
\vspace{-0.2cm}
\item
Relatively fewer {\it core-dominated} RLQs show associated NALs
than lobe-dominated RLQs do. 
\vspace{-0.2cm}
\item
The strongest associated NALs ($\gsim$ 3\AA) are mostly
in RLQs.
\vspace{-0.2cm}
\item
The data are consistent with the associated absorption being 
stronger in more inclined objects where reddening effects may play 
a larger role. Do these effects dominate radiation pressure effects?
\vspace{-0.2cm}
\item
About 8\% (5/66) of the RLQs show strong (EW $>$3\AA) absorption 
at relatively low velocities, which is not seen in the RQQs here.
While RQQs can accelerate the central outflow to high velocities (in BALs), 
the RLQs are perhaps not capable thereof to the same extent. This is 
consistent with predictions of disk outflow models.
Are RLQ NALs the low-velocity equivalents to the BAL phenomenon 
seen mostly in RQQs?

\end{itemize}
\begin{table}[h]
\vspace{-0.30cm}
\caption{\bf Frequency of Absorbed Objects and of C\,{\small \bf IV} NALs}
\begin{center}
\begin{tabular}{lrccc}
\vspace{-0.1cm}
\\[-18pt]
 \tableline\\[-8pt]
{\bf       }&{\bf  }&{\bf Absorbed QSOs} &{\bf Velocity $\leq$}    &{\bf 5000 $<$ Velocity} \\
{\bf Sample}&{\bf N}&{\bf (All Abs.\,Vel's)}&{\bf ~~5000 km s$^{-1}$}&{\bf ~~15000 km s$^{-1}$} \\ 
\tableline \\[-8pt]

All QSOs      & 114 & 66 = 58\% & 41 = 36\% & 36 = 32\% \\
RQQs          &  48 & 25 = 52\% & 14 = 29\% & 14 = 29\% \\
RLQs          &  66 & 41 = 62\% & 27 = 41\% & 22 = 33\% \\
CDQs          &  20 & 12 = 60\% & ~7 = 35\% & ~7 = 35\% \\
LDQs          &  46 & 29 = 63\% & 20 = 43\% & 15 = 33\% \\[3pt]
\hline \\[-8pt]
  \multicolumn{5}{c}{\bf Absorber Frequency Among Absorbed Quasars Only} \\[3pt]
\hline \\[-8pt]
LDQs          & $-$ &  29/66 = 44\% &  20/41 = 49\% &  15/36 = 42\% \\
CDQs          & $-$ &  12/66 = 18\% &  ~7/41 = 17\% &  ~7/36 = 19\% \\[3pt]
\hline
\tableline
\end{tabular}
\end{center}
\vspace{-0.95cm}
\end{table}
\noindent
\section{Sample, Data, and Measurements}
\vspace{-0.25cm}
The sample of 114 quasars (66 radio-loud and 48 radio-quiet) was selected 
for a study of the emission lines (Vestergaard 2000; Vestergaard, Wilkes, \& 
Barthel 2000; Vestergaard \et 2001). 
The spectra cover a minimum range from $\sim$1000 to $\sim$2100\,\AA\ with 
a spectral resolution of $\sim$\,5\,\AA{} or better.  High quality VLA radio maps 
at 1.4, 5, and 15\,GHz are available for most of the RLQs (Barthel \et 1988; 
Lonsdale, Barthel, \& Miley 1993; Barthel, Vestergaard, \& Lonsdale 2000).
Vestergaard \et (2001) provide details on all the data.

The RLQs are further subclassified as lobe-dominated quasars, 
LDQs (i.e., R$_{\rm 5 GHz}$ = S$_{\rm 5 GHz, core}$/S$_{\rm 5 GHz, total} <$ 0.5),
or as core-dominated quasars, CDQs (R$_{\rm 5 GHz} \geq$0.5). 
The quasars were selected such that the RLQs and RQQs have similar 
$z$ and luminosity, M$_{\rm V}$, distributions (e.g., Fig.\,1).

A cosmology of 
H$_0$ = 50 km\,s$^{-1}$\,Mpc$^{-1}$ and q$_0$ = 0 is used throughout.
\begin{figure}
\plotfiddle{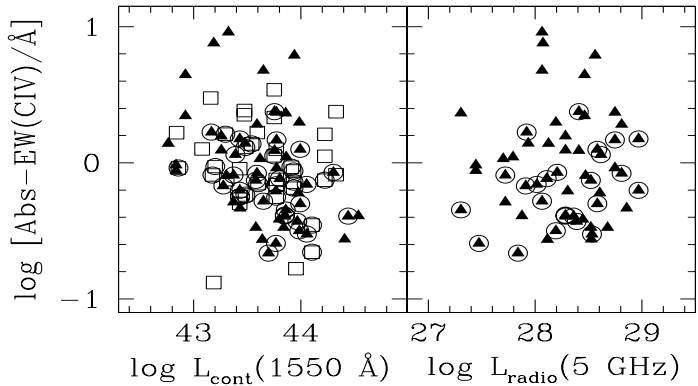}{4.4cm}{-90}{35}{30}{-140}{165}
\caption{The absorption strength versus continuum luminosities.  }
\vspace{-0.3cm}
\end{figure}
\vspace{-0.15cm}
\subsection{The Measurements}
\vspace{-0.15cm}
The \civ\ emission line profile was reproduced with a smooth fit [see
Vestergaard \et (2001) for details]. This smooth profile fit acts as the
local ``continuum'' for the absorption lines. No Galactic extinction
correction is performed on the spectra. 
The measured absorption lines were defined as {C\,{\sc iv} doublet absorption
based on the following criteria 
($\Delta \lambda_{\rm doublet}$\,=\,1550.77\,$-$\,1548.20\AA{}\,=\,2.57\AA):
\begin{itemize}
\vspace{-0.15cm}
\item
The observed equivalent width, EW$_{\rm obs}$, $\geq$ 0.5\AA{} of each system/blend
\vspace{-0.28cm}
\item
The measured 
$\Delta \lambda_{\rm separation,rest} = \Delta \lambda_{\rm doublet} \pm \half$ 
resolution element
\vspace{-0.28cm}
\item
Rest EW doublet ratio: 0.8 $-$ 2.2 (allows for blending/resolution effects)
\vspace{-0.28cm}
\item
The doublet FWHMs match within the resolution ($\sim$200$-$300 \kms)
\vspace{-0.28cm}
\item
Rest EW of each transition $\geq 3 \sigma$ detection limit
\vspace{-0.40cm}
\end{itemize}
\section{Results and Brief Discussion \label{results} }
\vspace{-0.15cm}
The results are presented in Table\,1 and in the figures. RLQs are shown 
as triangles, while RQQs are shown as open squares. Encircled symbols are 
high velocity ($>$5000 \kms) absorbers.

The quasars with NALs (Fig.\,1) have an average UV slope, 
$<\alpha_{\rm UV}>$\,=\,$-$1.54 (median\,=\,$-$1.48), while unabsorbed quasars 
have $<\alpha_{\rm UV}>$\,=\,$-$1.99 (median\,=  $-$1.89). However, a K-S test 
shows no statistically significant differences at the 99.95\% confidence level.
Both groups have $<$M$_{\rm V}$$>$$\,\approx\,-$27.9 mag.
The strength (equivalent width, EW) of the \civ\ NALs correlates strongly with 
$\alpha_{\rm UV}$ (Spearman's rank, r\,=\,0.43) with a P\,$<$0.1\% probability 
of occuring by chance (Fig.\,2, left). A slightly weaker correlation (r\,=\,0.21, 
P\,=\,3.95\%) exists with quasar luminosity (Fig.\,2, right) such that the EW 
tends to decrease in brighter objects. A stronger trend is seen with UV continuum 
luminosity (r\,=\,$-$0.32, P\,=\,0.19\%; Fig.\,3, left).  This is contrary to 
the na\"{\i}ve expectation (\S\,1) that brighter, bluer objects more easily 
generate stronger disk-winds. As will be clear later, what is encountered 
is a complication due most likely to inclination and/or reddening effects.
No relation is seen with radio luminosity (Fig.\,3, right).

\begin{figure}[t]
\plotfiddle{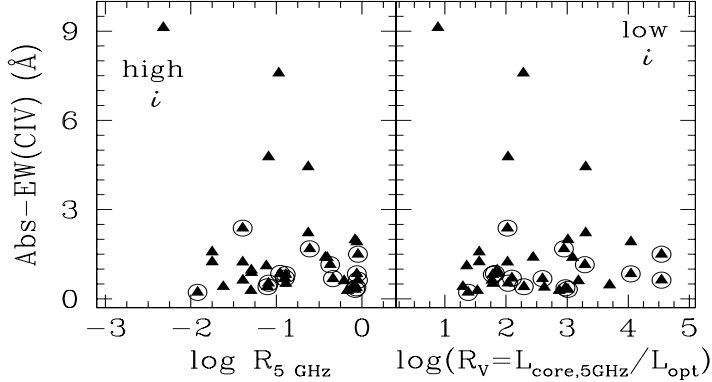}{4.35cm}{-90}{35}{30}{-140}{162}
\caption{The absorber EW versus source inclination ($i$) estimators.}
\vspace{-0.2cm}
\end{figure}
\begin{figure}[t]
\plotfiddle{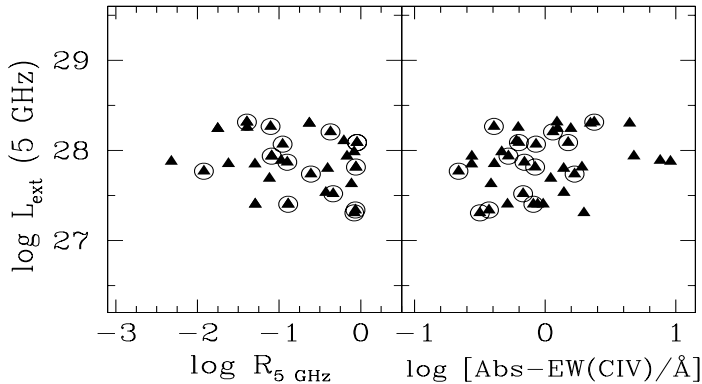}{4.4cm}{-90}{35}{30}{-140}{165}
\caption{
The EW inclination dependence is {\it not} due to L$_{\rm ext}$ biases.
}
\end{figure}
\begin{figure}[h]
\plotfiddle{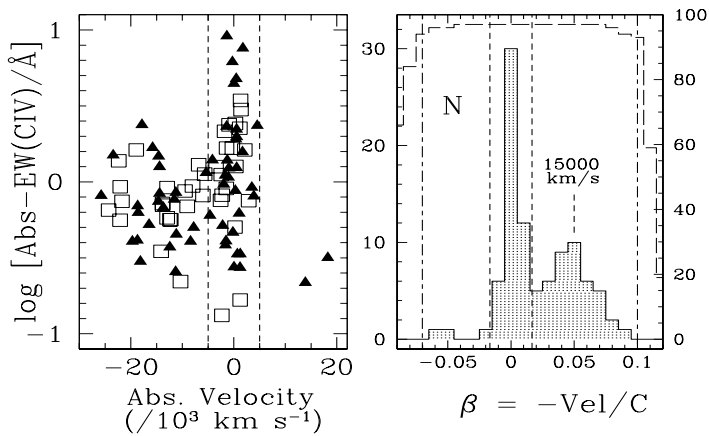}{4.8cm}{-90}{35}{30}{-140}{167}
\caption{The distribution of absorber strength on the absorber velocity
relative to the emission redshift (left). {\it Right:} The number distribution
of the absorbers (shaded; left axis) compared with the spectra measured
(dashed curve; right axis) show both peaks are real; see text.}
\vspace{-0.4cm}
\end{figure}

Table\,1 lists the frequency of narrow \civ\ absorbers among various quasar 
subgroups.  The main results were listed in the summary. Not only do LDQs have 
a higher NAL frequency than CDQs, but they also tend to be more strongly absorbed 
(Fig.\,4; also e.g., Foltz \et 1988; Barthel, Tytler, \& Vestergaard 1997, who 
used almost the same RLQ sample studied here).  LDQs are believed to be 
intrinsically similar to CDQs, just viewed at a higher inclination, $i$, of the 
radio axis relative to our line of sight.  In this study the $i$-dependence of 
the NAL strengths is tested on a more well-suited sub-sample of the RLQs, which 
is important if the NALs are associated with the quasars. 
To avoid possible selection biases the LDQs and CDQs were selected to cover the 
same range in L$_{\rm ext}$ (Fig.\,5; Vestergaard \et 2000). Also, no L$_{\rm ext}$ 
dependence is seen for the EWs.  Both $\log R_{\rm 5GHz}$ and $\log R_{\rm V}$ 
estimate $i$ (e.g., Wills \& Brotherton 1995). Also for this subsample are the 
strongest NALs seen in LDQs and the strength increases with $i$ (Fig.\,4).  If 
indeed LDQs are highly inclined, then the correlations of NAL EW with 
$\alpha_{\rm UV}$, M$_{\rm V}$, L$_{\rm cont}$, $\log R_{\rm 5GHz}$, and 
$\log R_{\rm V}$ (Figs.\,2, 3, 4) can be explained as a combination of inclination 
and reddening effects, which thus seem to dominate possible radiation pressure 
effects on disk outflows.  It is worth noting that Ganguly \et (2001) do {\it not} 
find an enhanced frequency of \civ\ NALs for the few LDQs among 59, $z \la$1.2 
quasars from the {\it HST} Quasar Absorption Line Key Project. Furthermore, their 
measured NAL strengths are significantly lower indicating a clear redshift 
evolution of the NALs.  And Richards \et (2001) find a small excess of 
high-velocity NALs in CDQs relative to LDQs.  
These issues will be addressed in forthcoming work.

The strongest NALs are furthermore {\it associated} with the quasars. Fig.\,6 
shows the distribution of EW with absorber velocity relative to the quasar 
restframe.  With exception of the usual five strongest RLQ NALs, there is a 
similar velocity distribution for RLQs and RQQs (Fig.\,6, left). Fig.\,6 (right) 
shows the distribution of the $\beta$ parameter [$\beta = (r^2 - 1)/(r^2 + 1)$, 
$r = (1 + z_{\rm em})/(1 + z_{\rm abs})$; e.g., Peterson 1997]. The enhancement 
of NALs within 5000 \kms\ (between the vertical, short-dashed lines) is clear. 
The long-dashed curve shows the number of spectra available for measurement at a 
given $\beta$ bin.  The vertical, dot-dashed lines denote the range where at least
95\% (92/96) of the spectra are present. The high fraction of spectra across the 
entire second bump shows that the 15,000\,\kms\ peak is real.  The increased EW around 
18,000 \kms\ (Fig.\,6, left) is also quite possibly real. These two peaks are 
intringuingly close to the terminal velocities typically seen in BALs 
($\sim$20,000\,\kms); this may potentially be important.

The fact that RQQs tend to show modest NAL EWs, while RLQs have quite strong NALs
is consistent with disk-wind models (e.g., Murray \& Chiang 1995) which predict
that RLQs are not capable of accelerating the high density outflows to relativistic
velocities because the stronger X-ray flux strips the electrons off the atoms,
thereby decreasing the radiation pressure on the outflowing gas. The moderately 
strong, high-velocity NALs seen in the RLQs ($\sim$18,000 \kms\ peak, Fig.\,6, 
left) adds an interesting twist to this scenario. Perhaps the NALs are a separate 
subset of the equatorial absorbers as they are seen independently of radio-type. 
This and the two peaks of high-velocity absorbers will be further addressed by 
Vestergaard (2001).

\acknowledgements
I thank the conference organizers and the participants for a very
enlightening and inspiring workshop.
%

\end{document}